\documentclass{ws-procs9x6}
\DeclareMathAccent{\pol}{\mathord}{letters}{"7E}
\newcommand{\be}{\begin{equation}}
\newcommand{\ee}{\end{equation}}
\newcommand{\bdm}{\begin{displaymath}}
\newcommand{\edm}{\end{displaymath}}
\newcommand{\bea}{\begin{eqnarray}}
\newcommand{\eea}{\end{eqnarray}}
\newcommand{\no}{\nonumber \\}

\begin{document}

\title{Opening Remarks at Chiral Dynamics 2006: Experimental Tests of Chiral Symmetry Breaking}

\author{A.M.Bernstein \footnote{Introductory talk at the Chiral Dynamics 
Workshop 2006, Duke University, Sept. 2006.  E-mail: bernstei@lns.mit.edu}}

\address{Department of Physics, LNS, MIT,\\
Cambridge, MA 02139, USA}

\begin{abstract}
A physical  introduction to the basics of chiral dynamics  is presented. Emphasis is placed on  experimental tests. These have generally demonstrated a strong confirmation of the predictions of chiral perturbation theory, a low energy effective field theory of QCD. Special attention is paid to a few cases where discrepancies exist, requiring  further work. Some desirable future tests are also recommended. 
\end{abstract}


\bodymatter

\section{ Brief History of this Workshop Series and an Introduction to Chiral Dynamics}
\label{sec:intro}

This workshop is the fifth of a series dedicated to  Chiral Dynamics: Theory and Experiment that Barry Holstein and I started at MIT in 1994\cite{CD1994}. At that time, the theory was generally far ahead of experiments, and we decided that it was important to start serious discussions in which the experimental aspects of the field would be treated on an equal footing with the theory. We also decided that it should be a workshop  (not a conference) allowing plenty of time for active discussion. The format of having plenary talks in the mornings and working groups in the afternoons was established. Our goal was to  evaluate what had happened in the field in the last few years and to discuss where we should  go in the near future. In the early period of the workshops we decided to have them every three years and to alternate between Europe and the U.S. We  have now had workshops in  Mainz  (1997\cite{CD1997}), Jefferson Lab ( 2000\cite{CD2000}), Bonn (2003\cite{CD2003}), and now at Duke. I am very proud that this series has become an important measure of the progress of this field.

As the first speaker of this workshop I thought that it was important to introduce some of the terminology and concepts of chiral dynamics and the experiments that test its  theoretical predictions. As is well known the coupling between quarks in QCD increases at low energies (for an introduction to QCD and chiral dynamics see e.g.\cite{DGH}) which leads to confinement. This has the consequence that  normal perturbation theory does not work at low energies. However there is a low energy effective theory  known as chiral perturbation theory(ChPT)\cite{BM-review,Bernard-review,gl-su2}. The QCD Lagrangian can be written as a sum of two terms, $L_{0}$ which is independent of the light quark masses (up, down, strange) and $L_{m}$ which contains the masses of the  three light quarks. Consider the chiral limit in which the three  light quark masses $m_{q} \rightarrow 0$.  The solutions to the Dirac equation for massless quarks  have a definite chirality or equivalently  helicity = $\hat{\sigma}\cdot \hat{p} =\pm 1$. The terminology is that when the quark spin  $\hat{\sigma}$ is parallel (anti-parallel) to the momentum vector $\hat{p}$, the quark is labeled as right(left) handed. For massless quarks, the left and right handed solutions are independent and this is known as chiral symmetry. Another language to express this is that there are separate conservation laws for vector (left +right) currents  and axial vector(left- right) currents\cite{DGH}. As is well known, the vector current is conserved while the axial vector current is conserved only in the chiral limit (i.e. $m_{q} \rightarrow 0$) and slightly non-conserved in the real world. This is one of the approximate symmetries of QCD which was earlier known as current algebra/PCAC(partially conserved axial currents)\cite{DGH}. 

Despite the fact that the light quark mass independent part of the QCD Lagrangian  $L_{0}$, has chiral symmetry, matter does not seem to obey the rules. The chiral symmetry is expected to show up by the parity doubling of all hadronic states, i.e., the proton with $j^{p} =1/2^{+}$ would have a $1/2^{-}$ partner(the Wigner-Weyl manifestation of the symmetry). Clearly this is not the case. This indicates that the symmetry is spontaneously hidden (often  stated as spontaneously broken) and is manifested in the Nambu-Goldstone  mode; the parity doubling occurs through  the appearance of a massless pseudo scalar ($0^{-}$) meson. The opposite parity partner of the proton is a proton and a "massless pion". 

There are profound  consequences of spontaneous chiral symmetry hiding  which are subject to experimental tests. In the SU(2) version of the picture the up and down quark masses are considered small and the three $\pi$ mesons are the Nambu-Goldstone Bosons. In the SU(3) version the up, down, and strange quark masses are considered small and there are eight Nambu- Goldstone Bosons ($\pi^{\pm,0}, \eta, K^{0,\pm}, \bar{K^{0}})$. Below the chiral symmetry breaking scale $\Lambda_{x} \simeq $1 GeV all of the lightest hadrons are Nambu-Goldstone Bosons with the pions $\simeq$ 140 MeV, the $ \eta \simeq$ 547 MeV and the Kaons $\simeq$ 496 MeV\cite{PDB}. Clearly these masses are not zero, due to the explicit chiral symmetry breaking term $L_{m}$ in the QCD Lagrangian. The lowest order estimate for the pion mass $m_{\pi}^2 = B_{0} (m_{u} + m_{d})$ (Gell-Mann, Oakes, Renner relation) where $B_{0}$  is proportional to the scalar quark vacuum  condensate $< 0 \mid \bar{q} q \mid 0>$ where q represents up or down quarks. This is an order parameter of QCD which mixes left and right handed states. From this formula one can see that $m_{\pi} \rightarrow 0$  in the chiral limit $m_{q} \rightarrow 0$. This formula for $m_{\pi}$ represents the strong but not the electromagnetic interaction contribution. The experimentally observed mass difference $m_{\pi^{\pm}} -m_{\pi^{0}} = 4.59 $ MeV\cite{PDB} is almost purely electromagnetic in origin. Since the lightest hadron is the  $\pi^{0}$ meson its primary decay mode is $\pi^{0} \rightarrow \gamma  \gamma$. This allows us to perform a precision test of the predictions of the QCD axial anomaly which is the principal mechanism for this decay\cite{DGH} (see Sec.\ref{sec:Axial}). 

The formulas for the masses of the K and $\eta$ mesons  contain  the strange quark mass $m_{s}$. From the masses of these mesons (subtracting the electromagnetic contribution) the ratios of the light quark masses can be accurately obtained. For example the ratio $m_{u}/m_{d} = 0.553 \pm 0.043$ has been obtained\cite{L:masses}. The fact that all of the up and down quark  masses differ by almost a factor of two  means that there is strong isospin (SU(2)) breaking in addition to electromagnetic effects \cite{W:mass}. However since both of these masses are  small the magnitude of this will be  typically $\leq$ 1\% \cite{Meissner:IS} [Meissner]\footnote{references to talks at this meeting will be presented in brackets}. The large strange quark mass compared to the up and down quark comes from the fact that the kaon and $\eta$ are much more massive than the pion. Physically this means that the most accurate tests of chiral dynamics are in the pion sector.   
The pseudoscalar $\eta^{'}$ meson with a mass of $ \simeq$ 958 MeV is very interesting. Its large mass is due to the QCD axial anomaly\cite{DGH} and therefore  has a large component which does not vanish in the chiral limit. On the other hand, in the large $N_{c}$(number of colors) limit it would be the ninth Nambu-Goldstone Boson. 

To obtain absolute values of the quark masses takes additional assumptions (e.g. QCD sum rules and/or input data\cite{Narison} or, more recently, lattice calculations. Although the situation is improving there are much larger errors then in the mass ratios. For a summary of the literature see the particle data book\cite{PDB} where the ranges for 
$m_{u}$ from 1.5 to 3.0 MeV, $m_{d}$ from 3 to 7 MeV, and $m_{s} = 95 \pm 25$ MeV are given\footnote{In QCD  the coupling constant $\alpha_{s}$  and quark masses are scale and renormalization scheme dependent (see e.g. the QCD section of the particle data book\cite{PDB}). It is customary to use the "modified minimum subtraction scheme" $\bar{MS}$,  to quote the value of $\alpha_{S}$ at the mass of the Z gauge boson, and the quark masses at a scale of 2 GeV.}. 

The effective field theory  that utilizes the concepts of spontaneously hidden chiral symmetry is called chiral perturbation theory (ChPT)\footnote{For an introduction to ChPT  see\cite{BM-review,DGH,L:intro}  and for more complete reviews  see\cite{Bernard-review, Scherer-review,L:reviews}.} This is an effective (low energy) theory of QCD in which  
the quark and gluon fields are replaced by a set of  fields U(x) describing the degrees of freedom of the observed hadrons.
For the Nambu-Goldstone Boson sector this is usually taken to be of the non-linear  exponential form $U(x) = exp[i \phi(x)/F_{\pi}]$ where $F_{\pi}$ is the pion decay constant $\simeq$ 92 MeV and $\phi$ represents the Nambu-Goldstone fields, a 
$2\!\times\!2$ matrix for the pion fields if we assume only that the up and down quarks are active, and a  $3\!\times\!3$ matrix representing  pion, $\eta$, and Kaon fields when we take the strange quark into account. 
The Lagrangian of QCD is replaced by an
effective Lagrangian, which only involves the field U$(x)$, and  even powers of its derivatives
$  L_{QCD} \rightarrow  L_{eff}(U,\partial{U},\partial^2{U} ,\ldots) = L_{eff}^{2} +
L_{eff}^{4} +  L_{eff}^{6} +\ldots $ where the superscript n on $L_{eff}^{n}$ represents the number of derivatives. The form of the terms are fully determined by the requirements of chiral symmetry. However the magnitudes of the terms are not determined by the symmetries and must be determined empirically or on the lattice. These  are in reasonable agreement with model estimates\cite{gl-su2,BM-review,Bernard-review}which shows that the physics is understood. For the SU(2) case the lowest (n=2) term is: 
%\begin{displaymath}\label{eff3}
$L_{eff} ^{2} = (F_{\pi}^{2}/4) [ \mbox{tr} 
\{\partial_\mu U^+ \partial^\mu U  \} +  
m_{\pi}^{2} \mbox{tr}  \{U + U^{+} \}]$
%\end{displaymath}
which contains the well known pion decay constant and mass. The  derivative  term predicts that Nambu-Goldstone Bosons are emitted and absorbed in p waves and have  no interaction as the momentum $\rightarrow$ 0 in accordance with Goldstone's theorem. The mass term explicitly breaks chiral symmetry and causes a small interaction at zero momentum.  

ChPT represents a systematic expansion with definite counting rules which determine the contributions at the order which one chooses to work. The predictions are  expansions in the Nambu-Goldstone Boson masses and momenta. To converge they must be small compared to the chiral symmetry breaking scale $\Lambda_{x} = 4 \pi F_{\pi} \simeq $ 1 GeV. They must also be in a region below any resonances or branch cuts. In $\pi$N scattering he energies must be significantly below the $\Delta$ resonance unless  it is included as a dynamical degree of freedom in the calculations\cite{Hemmert-Delta} [Pascalutsa]. In their domain of validity they represent the predictions of QCD subject to the errors which are imposed by uncertainties in the low energy constants and by the  neglect of higher order terms. As such they are worthy of great experimental effort in order to check them. Any discrepancy which is significantly larger than the combined experimental and theoretical errors must be taken seriously! 

\section { Chiral Dynamics Phenomena and Experiments }
    Following this brief introduction to the basic ideas of spontaneous chiral symmetry hiding in QCD and ChPT the phenomena  that are associated with this subject and the associated experimental possibilities will be outlined. 
     \begin{itemize}
\item{ Nambu-Goldstone Bosons  at Low Energies\\
        interactions: $\pi-\pi, \pi-K, \pi-\eta, \pi-\eta^{'}, K-\eta,  \ldots $  \\
        properties: RMS radii, polarizabilities \\
       electromagnetic and hadronic decays: $ \pi^{0}, \eta, \eta^{'}  \rightarrow \gamma \gamma, \\
      \gamma \gamma \rightarrow \pi \pi, \eta \eta ,  \eta \rightarrow \pi \pi \gamma,  \eta \rightarrow 3 \pi, \ldots $ \\
       leptonic and semi-leptonic decays: $\pi, K  \rightarrow e \nu \gamma,  K^{+} \rightarrow \pi^{+} l \nu_{l} \ldots $ } 
\item{Nambu-Goldstone Boson-hadron  scattering: $\pi-N, K-N, \ldots$}
\item{photo and electro-production of Nambu-Goldstone Bosons:\\
 $\gamma^{*} N \rightarrow \pi N, K \Lambda, \gamma^{*} \pi \rightarrow \pi \pi \ldots $}
\item{Hadron structure  at low $Q^{2}$  \\
      Nucleon EM,axial, strange form factors:\\  ~
        ~ RMS radii, magnetic moments 
      \\ quadrupole amplitudes in  $\gamma^{*} N \rightarrow \Delta $\\
 Electric and magnetic polarizabilities, $\pi N -\sigma$ term}
\item{ long range part of N-N interaction.  nuclear physics at low energies, nuclear astrophysics}
\end{itemize}
This  incomplete list   shows  the broad range of phenomena included in chiral dynamics. In this short presentation it is only possible to discuss a small fraction of these topics. The most pristine testing grounds for chiral dynamics is in the Nambu-Goldstone Boson section and $\pi-\pi$ scattering is the best case, both theoretically and experimentally. However experiments are difficult since precision is vital and there are no free pion targets. The best experimental method has been to study the final state interactions in $K \rightarrow \pi \pi l \nu$ and more recently by the cusp in $K \rightarrow 3 \pi$. Since this subject was extensively covered in this workshop[Leutwyler, Goy-Lopez, Pelaez, Bedaque] I will  say only  that it is still  very interesting and evolving. More open is the study of $\pi-K$ interactions for which values of the scattering lengths have been extracted from higher energy data using dispersion theory\cite{Bachir:pi-K} based on data from the  KN $ \rightarrow \pi$ KN reaction. Lower energy data has been taken in the decay $D^{+} \rightarrow \pi^{-} K^{+} \mu^{+} \nu$ from the FOCUS collaboration at Fermilab\cite{FOCUS}, but have not yet been incorporated into this analysis.  As far as I know, there are no data on the interactions with the $\eta$ or $\eta^{'}$ ; the latter  is particularly interesting since it is a Nambu-Goldstone Boson in the large $N_{c}$ limit and it is not clear  how different its dynamics (e.g. scattering lengths)  will be  from, say, the $\eta$. 
 
      The polarizability of the pion has proven to be difficult to measure (for a summary see\cite{Filkov:2006} [Walcher, Kashevarov]). A recent experiment at Mainz\cite{Mainz-pion-pol} is in serious disagreement with ChPT predictions and the result extracted from  $e^{+}e^{-} \rightarrow \gamma \gamma \rightarrow \pi^{+}\pi^{-}$  experiments\cite{Gasser:pi-pol}.  In my view this is a potentially serious situation which urgently  needs additional theoretical and experimental effort. The kinematics of the Mainz experiment were chosen in order to maximize the sensitivity to the pion polarizability. The photon energy region of 537 to 817 MeV employed in this experiment is in a difficult region to analyze theoretically and in my view the small model error assigned to the pion polarizability needs to be justified. In addition, I think that an improvement in the accuracy of the $e^{+} e^{-}$ experiments is  needed since the sensitivity to the pion polarizability is not large. An excellent development in this field is that  a modern radiative Primakoff experiment ($\pi^{\pm} \gamma \rightarrow \pi^{\pm} \gamma$ has been performed by the Compass experiment at CERN. 
      
 The phenomena that are associated with the $\pi N$ system are at the heart of nuclear physics. The $\pi$ meson has a special role in the universe. The exchange of pions between nucleons (Yukawa interaction) is the long range part of the nucleon-nucleon potential, and governs low energy nuclear interactions and stellar formation. Indeed the effective theory of few nucleon systems[Nogga, Hahnhart] has become an integral part of chiral dynamics and of these workshops. The pion cloud which surrounds hadrons plays a major role in their structure, e.g. their form factors and polarizeabilities[McGovern, Miskimen]. The $\Delta$ resonance (the first excited state of the nucleon) plays a central role in $\pi N$ dynamics, a topic discussed below. 
 
The Nambu-Goldstone Boson- Fermion sector  is more  difficult theoretically (compared to the purely Nambu-Goldstone Boson physics) but parts of it, e.g. $\pi N$ physics, including low energy  electromagnetic meson production, are much more easily  accessible experimentally which has lead to  extensive and accurate experimental data. This does not  mean that all problems are solved. For example there is an old open problem that was not even discussed at this workshop, namely the value of the $\pi N- \sigma$ term. Generally the interpretation of this leads to a greater than 20\% contribution of the strange quark to the nucleon mass. Yet, as we learn from parity violating electron scattering [Michaels], the strangeness magnetic moment is close to zero. There is no  simple connection between these quantities  since they are the expectation values of different operators. However, the fact that one is large and the other is small, needs explanation.  
  
\vspace{-0.5  cm}

\section{Pion-Nucleon Interactions and Electromagnetic Pion Production}
\label{piN}
In this and the next  section I shall give a few specific examples to illustrate some of the specific applications of chiral dynamics. From the large number of possibilities I have chosen ones that I have worked on, and are close to my heart. 
 
The $\pi N$ interaction in momentum space = $g_{\pi N} \vec{\sigma} \cdot \vec{p_{\pi}}$ where $\vec{\sigma}$ is the nucleon spin. In accordance with Goldstone's theorem, this interaction  $\rightarrow 0$ as the pion momentum $\rightarrow 0$. Furthermore $g_{\pi N}$ can be computed from the Goldberger-Treiman relation\cite{DGH} and chiral corrections\cite{Goity:GT}, and is accurate to the few \% level. The $\pi N$  interaction is very weak in the s wave and strong in the p wave which leads to the $\Delta$ resonance, the tensor force between nucleons, and to long range non-spherical virtual pionic contributions to hadronic structure. 
 For illustrative purposes consider the lowest order ChPT calculation $O(p^{2}$) for $a(\pi,h)$, the s wave $\pi$ hadron scattering length;  
  $a^{I}(\pi,h) = - \vec I_{\pi} \cdot \vec{I_{h} }m_{\pi}/(\Lambda_{x}  F_{\pi})$
  where  $\vec{I} = \vec{I_{\pi}} + \vec{I_{h}} $ is the total isospin, and $I_{\pi}$, and $I_{h}$ are the isospin of 
the pion and hadron respectively, $F_{\pi}$ is the pion decay constant, and $\Lambda_{x} = 4 \pi F_{\pi} \simeq $ 1 GeV is the chiral symmetry breaking scale\cite{W:pion-scat}. Note that $a(\pi,h)\rightarrow  0$ in the chiral limit $ m_{\pi} \rightarrow  0$ as it must  to obey Goldstone's theorem. Also note that  $a(\pi,h) \simeq 1/\Lambda_{x}  \simeq $ 0.1 fm, which is small compared to a typical strong interaction scattering length of $\simeq$ 1 fm. This small scattering length  is obtained from the explicit chiral symmetry breaking due to the finite quark masses. The predictions of ChPT for $\pi$N scattering lengths have been verified in detail in a beautiful series of experiments on pionic hydrogen and deuterium at PSI\cite{Gotta:atoms}; this includes the isospin breaking due to the difference in $m_{d} -m_{u}$ mentioned previously [Meissner].
   
Low energy electromagnetic production of Goldstone Bosons is as 
fundamental as Goldstone Boson scattering for two  
reasons: 1) the production amplitudes vanish in the chiral limit (as 
in scattering); and 2) the phase of the production amplitude is linked to 
scattering in the final state by unitarity or final state interaction (Fermi-
Watson) theorem suitably modified to take the up, down quark masses into account\cite{AB:FW}. 
First consider the low energy limit of the electric dipole $E_{0+}$ for s wave photo-
pion production\cite{BKM-photo}:
\begin{equation}
\begin{array}{rcl}
E_{0+}(\gamma p \rightarrow \pi^{0}p) & = &- D_{0} \mu ( 1 +
O(\mu)+..)\rightarrow 0\\
 E_{0+}(\gamma p \rightarrow \pi^{+}n) & = & \sqrt{2} D_{0}/(1+ \mu 
+...)^{3/2} \rightarrow \sqrt{2} D_{0}\\
\mu & = & m_{\pi}/M \rightarrow 0\\
D_{0} =e \cdot g_{\pi N}/8 \pi M& = & 24 \cdot 10^{-
3 }(1/m_{\pi})\\
\end{array}\label{eq:E0}
\end{equation}
where M is the nucleon mass and the right arrow denotes the chiral limit 
($m_{u}, m_{d},m_{\pi} \rightarrow 0$). Eq.~\ref{eq:E0} shows
that for neutral pion production the amplitude vanishes in the chiral limit. 
For charged pion 
production, there is a different low energy theorem\cite{BKM-photo}. Therefore the 
amplitude that is most sensitive to explicit chiral symmetry breaking is 
neutral pion production and most of the modern 
experiments have concentrated on this channel. In general, ChPT to one 
loop calculated in the heavy Fermion approximation, has been highly 
successful in calculating the observed cross sections and linearly polarized photon asymmetry\cite{BKM-photo}. 

\begin{figure}
\begin{center}
\epsfig{file=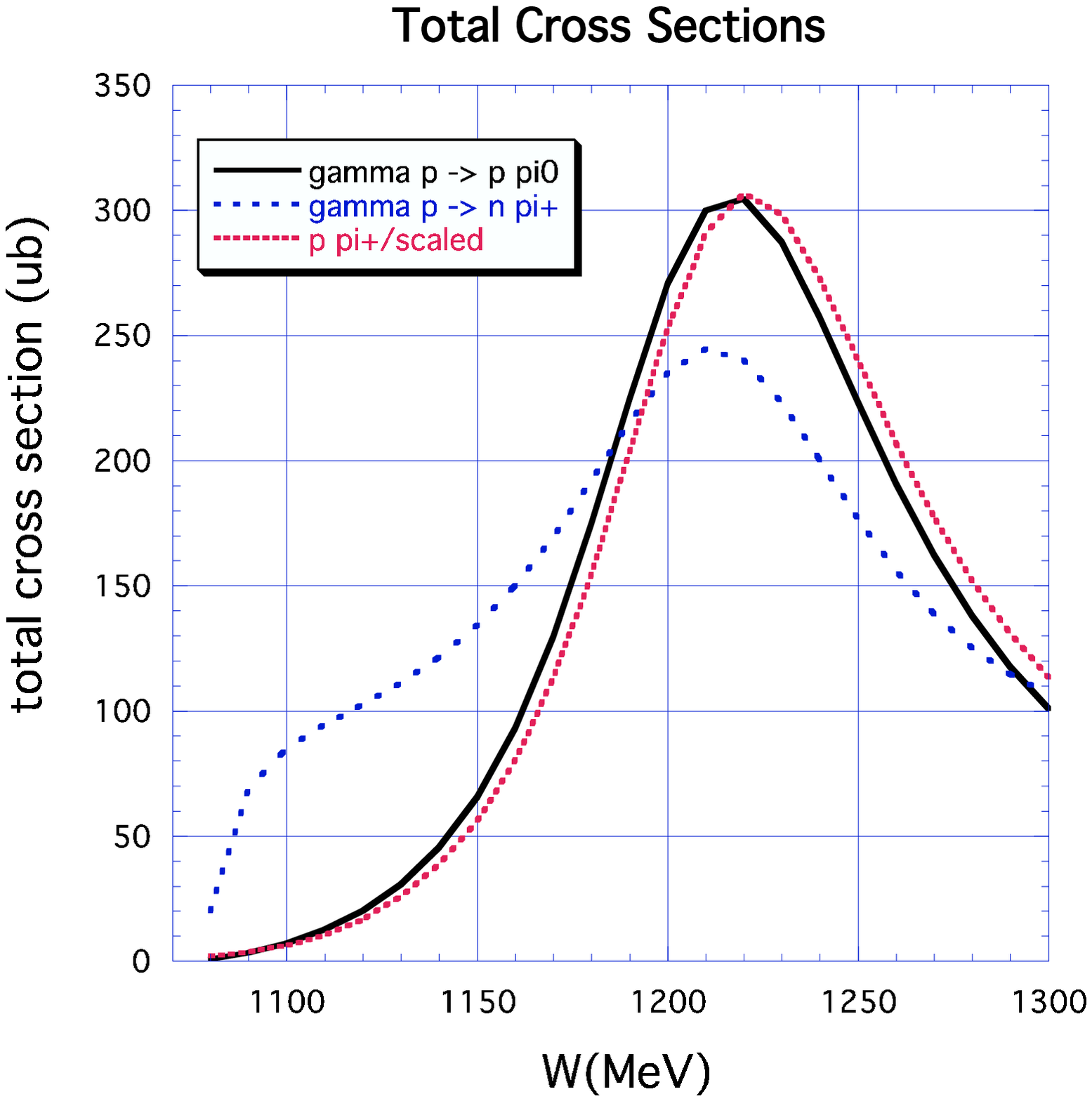,width=1.9in}
\epsfig{file=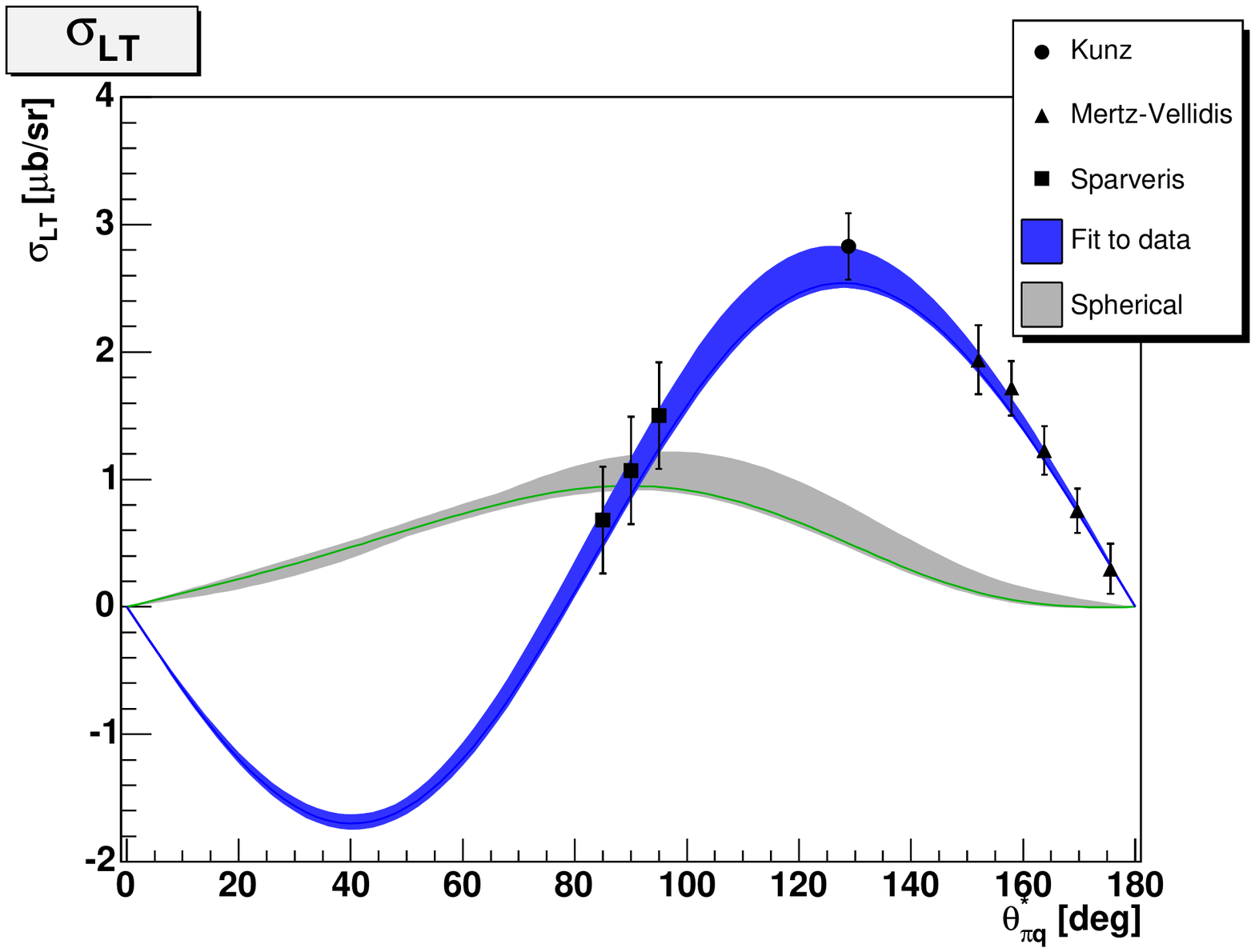, width=2.5in}
\end{center}
\caption{\label{fig:shape}
Left panel: The shape of the $\Delta$ resonance from fits to the total cross sections for $\pi^{+} p$(scaled solid curve) scattering and and for the $\gamma p \rightarrow \pi^{0} p$(short red dots), and $\gamma p \rightarrow  \pi^{+} n$ reactions(blue dotted curve which is larger at low W) versus W, the center of mass energy.the curves represent fits to the data\cite{SAID}. Right panel: $\sigma_{LT}$  from the $e p \rightarrow e^{'} \pi^{0} p$ reaction at $Q^{2} = 0.126 GeV^{2}$, W=1232 MeV(see text). The blue curve is  fit to the data\cite{nikos} and the relatively flat grey curve shows the calculation for the spherical case, i.e. when the quadrupole transition amplitudes are set to zero\cite{Costas, SOH}. }
\end{figure}

The application of these ideas to data from low energy $\pi N$ scattering and electromagnetic pion production from the nucleon is instructive. The left panel of Fig.\ref{fig:shape} shows the shape of the $\Delta$ resonance from fits to the total cross sections for $\pi^{+} p$ (scaled) scattering and and for the $\gamma p \rightarrow \pi^{0}p, \pi^{+} n$ reactions versus W (the center of mass energy)\cite{SAID}. All of the these reactions have a strong $\Delta$ resonance. The  $\pi^{+}p$ and $\gamma p \rightarrow \pi^{0}p$ reactions  have small cross sections near threshold and therefore clearly show the $\Delta$ resonance without any interference (the small shift between them is due to the mass difference of the $\Delta^{0}$ and $\Delta^{+}$). Indeed these  cross sections are text book example of an isolated resonance. Although not usually mentioned in text books  it is the combination of a strong resonance and a small cross section at threshold that produces this beautiful example (as predicted by chiral dynamics)!  In the case of the $\gamma p \rightarrow \pi^{+} n$ reaction there is  strong s wave production starting at threshold due to the Kroll-Ruderman low energy theorem (see Eq.\ref{eq:E0}). In this case the $\Delta$ resonance curve is superimposed on the strong s wave amplitude and looks quite different! 

The photo and electro-pion $\gamma^{*} p \rightarrow \Delta$ reactions have been extensively used to study non-spherical amplitudes (shape)  in the nucleon and $\Delta$ structure\cite{SOH}. This is studied by measuring the electric and Coulomb quadrupole amplitudes  (E2,C2) in the predominantly magnetic dipole, quark spin flip (M1) amplitude. At low $Q^{2}$ the non-spherical pion cloud is a major contributor to this (for a review see\cite{SOH}). Recently there have been chiral calculations of this process [Pascalutsa]. The right panel of Fig.\ref{fig:shape} shows our best estimate of the difference between the electro-excitation $\Delta$ for the spherical case(the relatively flat grey band) and the fit to the Bates data for the transverse-longitudinal  interference cross section $\sigma_{LT}$\cite{nikos} which shows the C2 magnitude which is primarily due to the pion cloud [Stave]\cite{Costas, SOH}. The evolution of the Coulomb quadrupole amplitude with $Q^{2}$ indicates that the quark models do not agree with experiment, but that models with pionic degrees of freedom do, demonstrating that the crucial ingredient in the non-spherical amplitude at long range is the pion cloud[Stave].  

\begin{figure}
\begin{center}
\epsfig{file=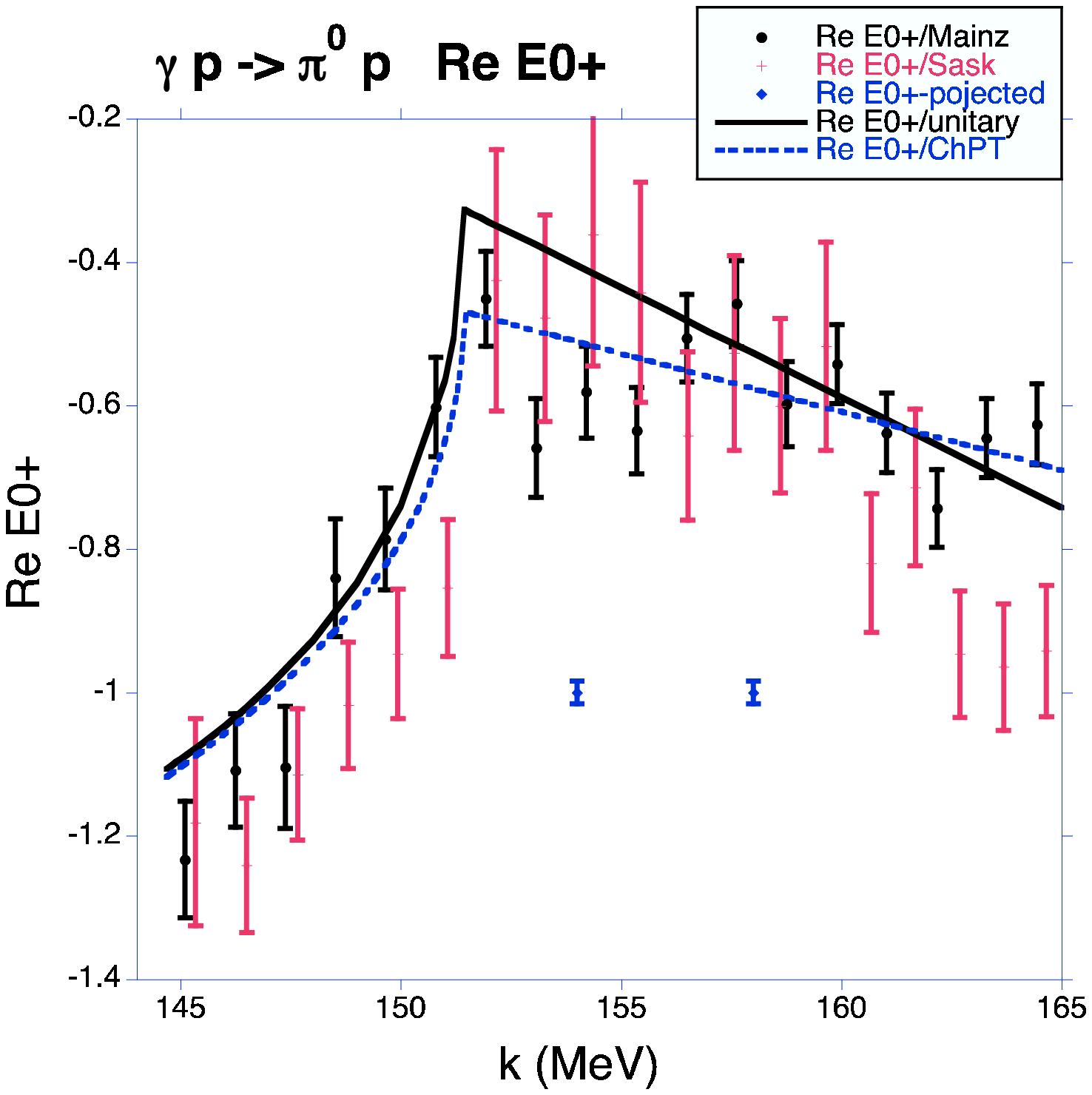, width=2in}
\epsfig{file=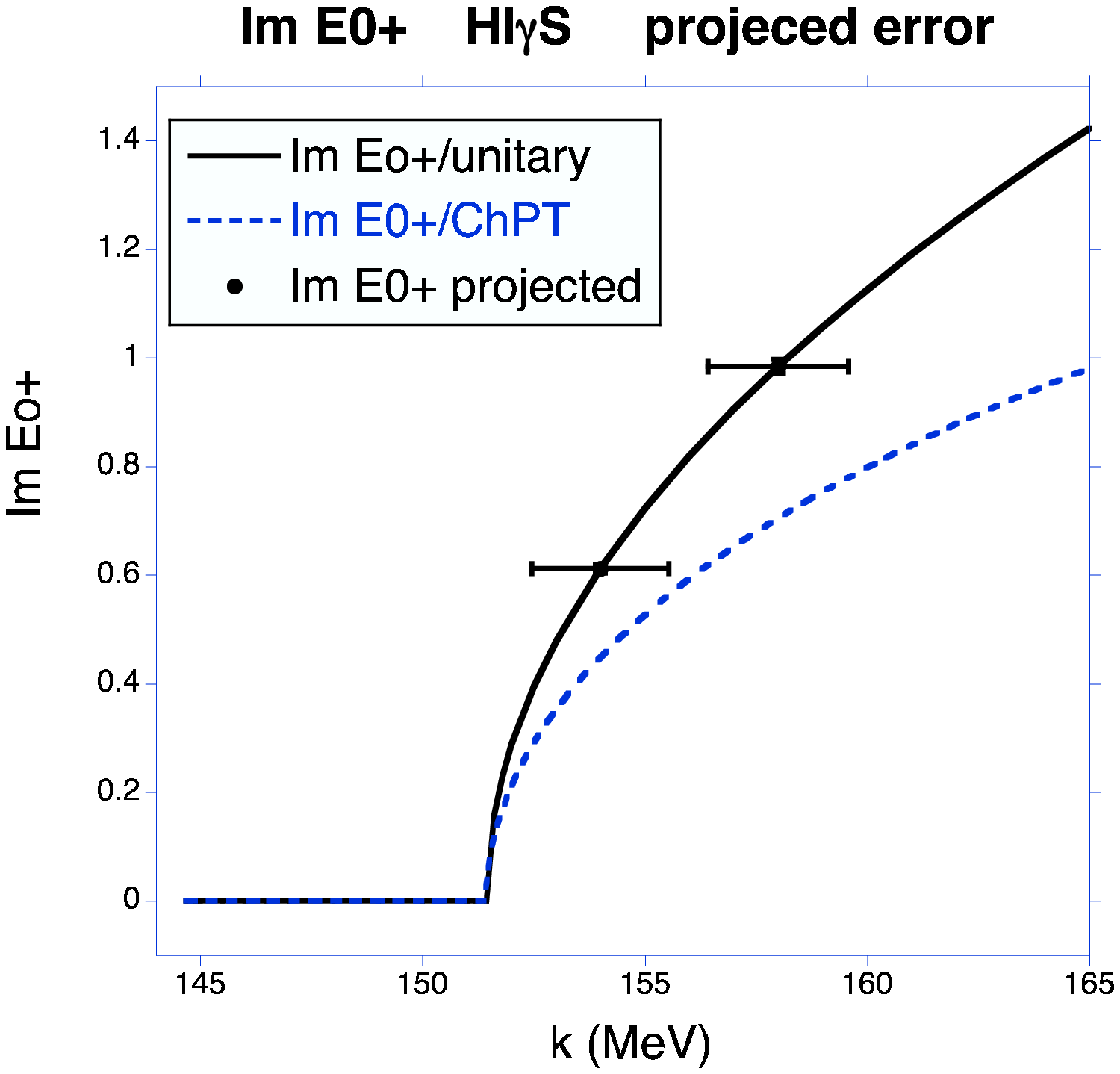,width=2in}
\end{center}
\caption{\label{fig:photo}
The $\gamma p \rightarrow \pi^{0} p$ Reaction. Left panel: Re$E_{0+}$ versus photon energy. The data points are from Mainz\cite{Schmidt} and Saskatoon\cite{Sask}. The curves are from ChPT\cite{BKM-photo} and a unitary fit to the data\cite{AB:FW}. The two projected points from HI$\gamma$S  are plotted at an  arbitrary value ($Re E_{0+}$ = -1)  to show the anticipated errors . Right panel:  Im $E_{0+}$ versus photon energy. The curves are the same as in the left panel and the projected HI$\gamma$S points are arbitrarily plotted on the unitary curve. There are no experimental points(see text).}
\end{figure}

A great deal of effort has gone into the study of the near threshold $\gamma p \rightarrow \pi^{0} p$ reaction experimentally at Mainz\cite{Schmidt} and Saskatoon\cite{Sask} and with ChPT calculations\cite{BKM-photo}. In addition we are planning to conduct future experiments at HI$\gamma$S, a new photon source being constructed at Duke[Weller]. These experiments will have full photon and target polarization and will be a  significant extension of the results we have at present.  The unpolarized cross sections were accurately measured  despite their small size and the results from Mainz and Saskatoon are in reasonable agreement. The p wave amplitudes tend to dominate even close to threshold. The real part of the  s wave electric dipole amplitude $ReE_{0+}$ is extracted from the data using the interference between s and p waves which goes as $cos(\theta_{\pi})$ in the differential cross section and leads to  larger errors. The results for  $ReE_{0+}$  versus  photon energy are plotted in the left panel of Fig. \ref{fig:photo}. There is reasonable agreement between the Mainz and Saskatoon points  as well as with  ChPT\cite{BKM-photo} and the unitary model calculations\cite{AB:FW}. The sharp downturn in $ReE_{0+}$ between the threshold at 144.7 MeV and the $\pi^{+}$ n threshold at 151.4 MeV is due to a unitary cusp caused by the interference  between the $\gamma p \rightarrow \pi^{0} p $ and $\gamma p \rightarrow \pi^{+} n $ channels. The magnitude of the cusp is $\beta= ReE_{0+}(\gamma p \rightarrow \pi^{+} n) \cdot a_{cex}(\pi^{+} n \rightarrow \pi^{0} p)$  which is measured to an accuracy of $\simeq$ 30\% from the data shown. The reason for this accuracy limitation is due to the fact that the  $ReE_{0+}$ is a sum of a smooth  and cusp functions and the smooth function is not known precisely\cite{AB:FW}. Therefore it is important to measure  $ImE_{0+}$ which starts from close to zero at the $\pi^{+}n$  threshold energy and rises rapidly as $\beta p_{\pi^{+}}$. This makes the extraction of $\beta$ as accurate as the measured asymmetry for $\pi^{0}$ photoproduction from a polarized target  normal to the reaction plane. The estimated error for such an experiment running at HI$\gamma$S for $\simeq$ 400 hours of anticipated operation of the accelerator is presented in the right panel of Fig.\ref{fig:photo}. This experiment, along with an independent measurement of the $\gamma p \rightarrow \pi^{+} n$ cross section will allow us to extract $\beta$ at the few \% level and measure the  charge exchange scattering length $a_{cex}(\pi^{+} n \rightarrow \pi^{0} p)$ for the first time. We will be able to  compare this to the measured value of $a_{cex}(\pi^{-} p \rightarrow \pi^{0} n)$\cite{Gotta:atoms} as an isospin conservation test. This illustrates the power of photopion reaction studies with transversely polarized targets to measure $\pi N$ phase shifts in  completely neutral  charge channels which are not accessible to pion beam experiments! This is potentially valuable to help  pin down experimentally the value of the $\pi N-\sigma$ term which has had a long, difficult measurement history.  

\begin{figure}
\begin{center}
\epsfig{file=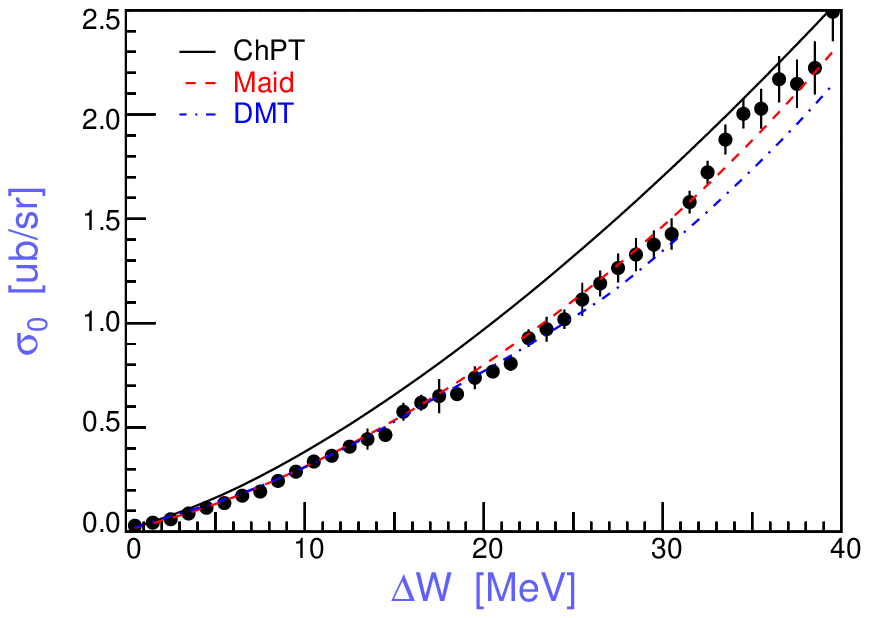, width=2.2in}
\epsfig{file=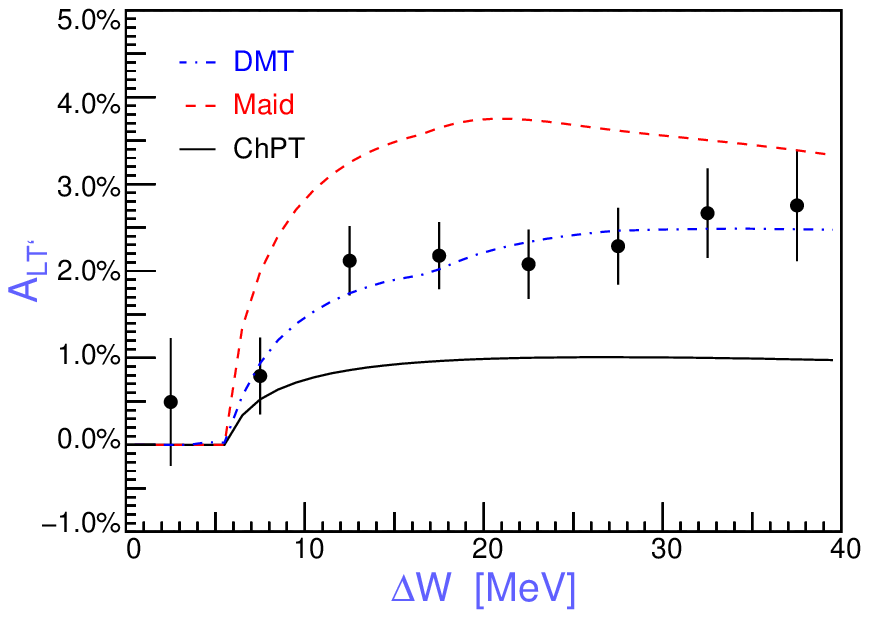,width=2.2in}
\end{center}
\caption{\label{fig:eepi}
Cross Section(Left panel) and $LT^{'}$ Asymmetrry (right panel)  for the $ep \rightarrow e^{'} \pi^{0}p$ Reaction at $Q^{2}  = 0.05 GeV^{2}$ versus dW, the center of mass energy above threshold\cite{Weis}. See text for discussion. } 
\end{figure}

ChPT has been extremely successful in predicting the cross sections and the linearly polarized photon asymmetry in the  $\gamma p \rightarrow \pi^{0} p$ reaction. However I would like to point out a significant discrepancy with the $e p \rightarrow e^{'} p \pi^{0}$ reaction data  at $Q^{2} =0.05 GeV^{2}$ taken at Mainz\cite{Weis}  shown in Fig.\ref{fig:eepi}. It can be seen that the ChPT calculations\cite{BKM-electro} do not agree with the data although the DMT dynamical model does\cite{DMT:threshold}. This discrepancy is a potentially serious problem which needs to be resolved!

\section{The $\pi^{0} \rightarrow \gamma \gamma$ Decay Width and the QCD Axial Anomaly}
\label{sec:Axial}

\begin{figure}
\begin{center}
\epsfig{file=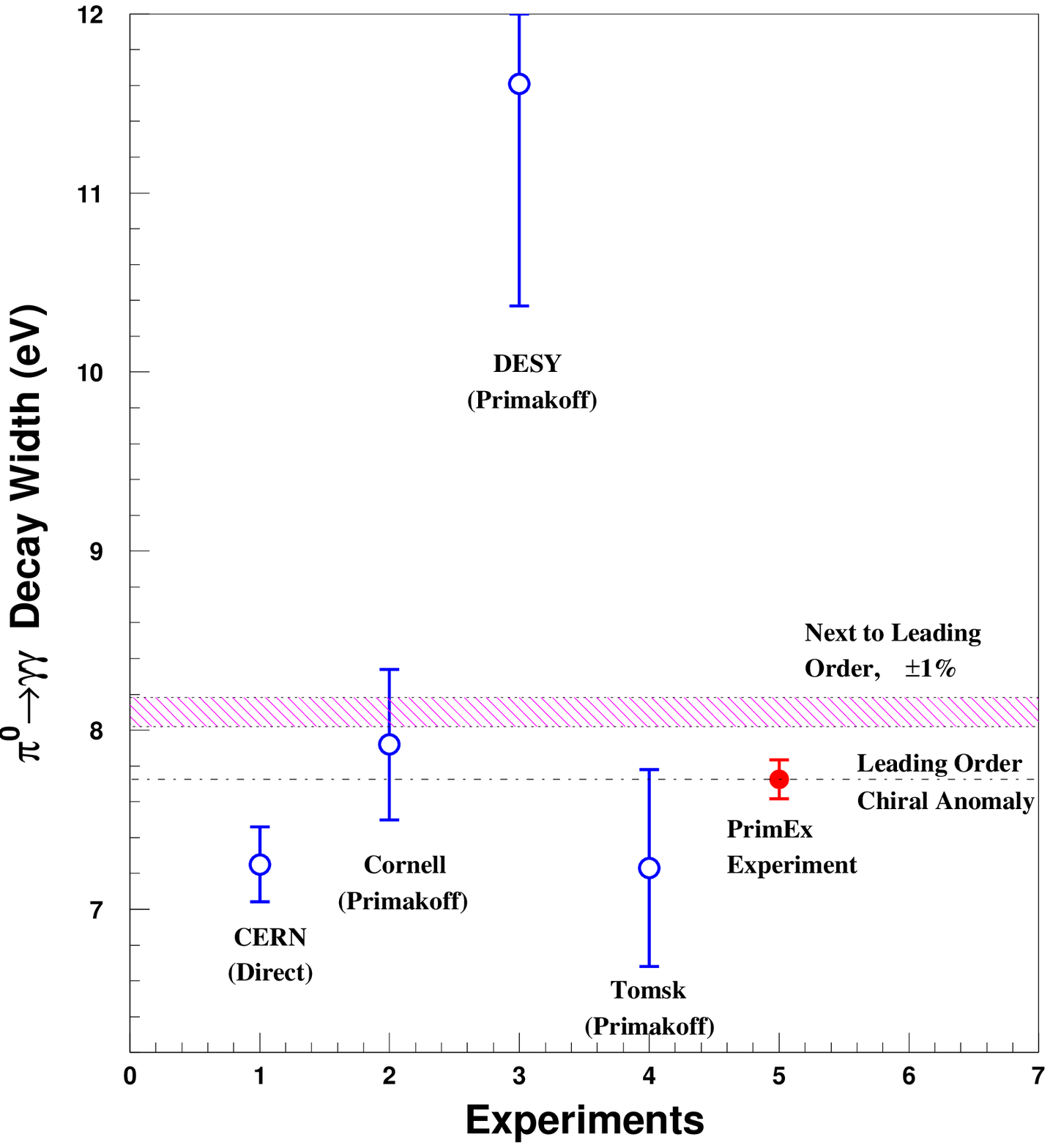, width=2in}
\epsfig{file=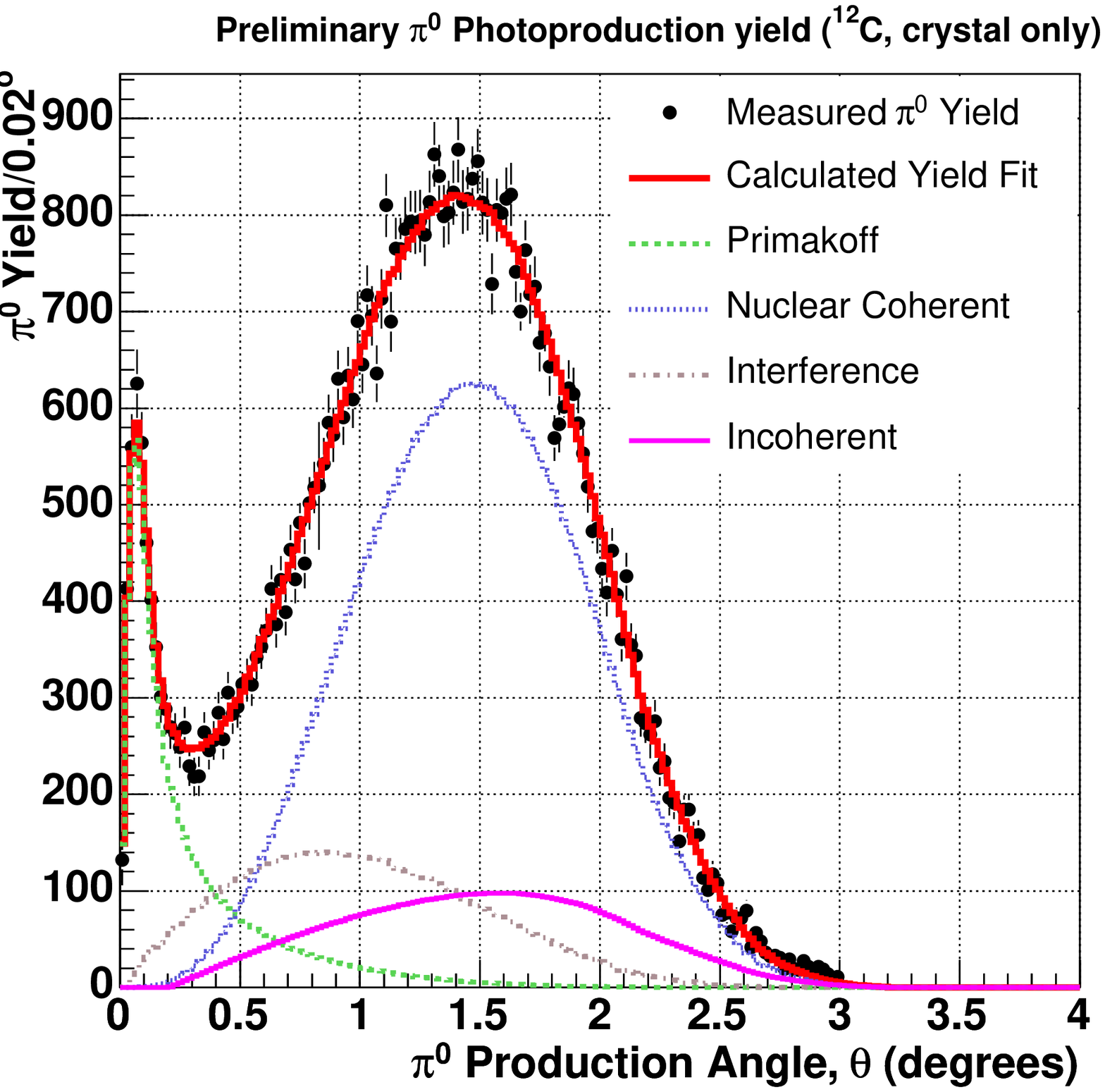,width=2.4in}
\end{center}
\caption{\label{fig:Prim}
Left panel:  the $\pi^{0} \rightarrow \gamma \gamma$ decay width in eV. These include the experimental points\cite{PDB}, the projected error of the Primex collaboration (arbitrarily plotted to agree with the predictions of the axial anomaly), and the next to leading order chiral correction\cite{pi0-decay}. Right panel: Preliminary yield and fit for the Primakoff effect on Carbon  versus pion angle with the individual contributions to the total yield exhibited (see text for discussion).}
\end{figure}

As the final special topic I would like to discuss a test of the axial anomaly by an accurate measurement of the $\pi^{0}$ lifetime. As was discussed in the introduction, due to the spontaneous breaking of chiral symmetry,  the $\pi^{0}$  is the lightest hadron and its primary decay mode is $\pi^{0} \rightarrow \gamma \gamma$. This decay rate is  exactly predicted in the chiral limit by the QCD axial anomaly. As is quoted in most textbooks on QCD (see e.g.\cite{DGH}) this prediction  is in agreement with the average  in the particle data book\cite{PDB} which has a  error $\simeq$ 10\%. However this oversimplifies the  experimental situation which is shown in Fig.\ref{fig:Prim}. In my opinion, almost all of the errors quoted in the literature are underestimates. This is indicated by the spread in the experimental values.  Also at issue are the chiral corrections to the decay rate. These have been worked out to next to leading order. They  primarily involve $\pi-\eta,\eta^{'}$ mixing which is isospin breaking and therefore proportional to $m_{d} -m_{u}$; they increase the predicted decay width by 4$\pm$ 1\% \cite{pi0-decay}. Another calculation based on QCD sum rules has also predicted a similar increase\cite{Ioffe}. 

This experiment has been performed by the Primex collaboration at JLab and the data analysis is in the final stages[McNulty]. The experiment measures the photo-production of $\pi^{0}$ mesons from C and Pb at an average  photon energy $\simeq$ 5.2 GeV and has the goal of achieving an accuracy of a few \%. It is the first Primakoff measurement to use a tagged photon beam. Prelilminary results are shown in  Fig.\ref{fig:Prim}. The large forward peak is due to the Primakoff effect which is the production of $\pi^{0}$'s in the Coulomb field of the target. The larger peak at a few degrees in C is due to  coherent nuclear production and there is a small quantum interference amplitude. The fits to these processes are shown. For the Pb target (not shown)  the nuclear coherent peak is small compared to the  Primakoff peak. This is due to final  state absorption causing the coherent nuclear cross section to scale $\simeq$ A, while the Primakoff cross section scales $\simeq Z^{2}$. Therefore the relative coherent to Primakoff peak decreases with heavier targets. As can be seen the preliminary data look  good and our collaboration expects to release preliminary lifetime results in the next half year.

\section{Conclusions}
 It is clear that chiral physics includes an impressive array of reactions and  particle properties. The general concepts and detailed calculations of ChPT are  generally verified for the  Nambu-Goldstone meson sector where they have been tested. The physics is  more complicated   when Fermions are also included. This sector is more closely related to the observable world, e.g. the properties of nucleons, and the formation of the stars. In this sector there are many areas of good agreement between ChPT and experiment but there are also a few outstanding problems

 Finally here is a short wish list that I would like to see granted by CD2009:
      \begin{itemize}
\item{further theoretical work on extracting the pion polarizability from he Mainz data, as well as further measurements  by different techniques to address the existing discrepancy}
\item{ further calculations and measurements of the $e p \rightarrow e^{'} p \pi^{0}$ reaction to resolve the existing discrepancy}
\item{accurate measurements of the decay $\pi^{0}, \eta, \eta^{'} \rightarrow \gamma \gamma$  decay widths}
\item{further isospin tests in the $\pi$N scattering  and $\gamma N \rightarrow \pi N$}
\item{progress on the $\pi N- \sigma$ term}
\item{ progress on the spin and $Q^{2}$  dependence of Nucleon polarizeabilities}
\item{progress on $\eta$ and $\eta^{'}$ decays\cite{COSY}}
\item{further calculations and experiments on the nature of $\eta^{'}$ Boson; what does it mean physically that this is a Nambu-Goldstone Boson in the large $N_{c}$ limit?}
\item{more data on the interactions between the heavier Nambu-Goldstone Bosons}
\end{itemize}

It was a pleasure for me to experience this workshop that was so stimulating and well organized. Personally I learned a  great deal in a collegial and enthusiastic atmosphere. I would like to thank that organizers and Duke University for their wonderful organization and hospitality. I would also like to thank V.R. Brown, M. Kohl, D. McNulty, and U. Meissner for their careful reading and constructive comments about this manuscript. This work has been supported in part by the U.S. Department of Energy under Grant No. DEFG02-94ER40818.

\bibliographystyle{ws-procs9x6}
\bibliography{Bernstein}

\begin{thebibliography}{10}

\bibitem{CD1994}
A.~M. Bernstein and B.~R. Holstein (eds.), {\em Chiral dynamics: Theory and
  experiment. Proceedings of the MIT Workshop, Cambridge, MA, USA, July 25-29,
  1994} (Springer Lecture notes in physics, Vol. 452).

\bibitem{CD1997}
A.~M. Bernstein, D.~Drechsel and T.~Walcher (eds.), {\em Chiral dynamics:
  Theory and experiment. Proceedings, Workshop, Mainz, Germany, September 1-5,
  1997} (Springer Lecture notes in physics, Vol. 513).

\bibitem{CD2000}
A.~M. Bernstein, J.~L. Goity and U.~G. Meissner (eds.), {\em Chiral dynamics:
  Theory and experiment. Proceedings, 3rd Workshop, Newport News, USA, July
  17-22, 2000} (World Scientific, 2001).

\bibitem{CD2003}
U.~G. Meissner, H.~H.W. and A.~Wirzba, Fourth Workshop on Chiral Dynamics -
  Chiral Dynamics 2003: Theory and Experiment, Bonn, Germany, Sept. 8-13,2003,
  hep-ph/0311212.

\bibitem{DGH}
J.~F. Donoghue, E.~Golowich and B.~R. Holstein, {\em Dynamics of the standard
  model} (Camb. Monogr. Part. Phys. Nucl. Phys. Cosmol., 1992).

\bibitem{BM-review}
V.~Bernard and U.-G. Meissner, Chiral perturbation theory,arXiv:hep-ph/0611231.

\bibitem{Bernard-review}
V.~Bernard, Chiral Perturbation Theory and Baryon
  Properties,arXiv:hep/ph0706.0312.

\bibitem{gl-su2}
J.~Gasser and H.~Leutwyler, Annals Phys.,158, 142 (1984) , Phys. Lett. B, 125,
  321 (1983), Phys. Lett. B,125, 325 (1983).

\bibitem{PDB}
W.~M. Yao {\em et~al.}, {\em J. Phys.} {\bf G33}, 1 (2006).

\bibitem{L:masses}
H.~Leutwyler, Nucl. Phys.B, Proc.Suppl.94,108(2001), J. Gasser and H.Leutwyler,
  Phys. Rept.87,77(1982).

\bibitem{W:mass}
S.~Weinberg, {\em Trans. New York Acad. Sci.} {\bf 38}, 185 (1977).

\bibitem{Meissner:IS}
B.~Kubis and U.-G. Meissner, Phys. Lett. B529, 69(2002); Fettes, Nadia and
  Meissner, Ulf-G, Phys. Rev.C63, 045201(2001).

\bibitem{Narison}
S.~Narison, {\em Phys. Rev.} {\bf D74}, p. 034013 (2006).

\bibitem{L:intro}
H.~Leutwyler, hep-ph/9409423 (article in \cite{CD1994}).

\bibitem{Scherer-review}
S.~Scherer, Adv. Nucl. Phys.,27(2003).

\bibitem{L:reviews}
H.~Leutwyler, hep-ph/0008124, hep-ph/9409422.

\bibitem{Hemmert-Delta}
T.~R. Hemmert, B.~R. Holstein and J.~Kambor, {\em J. Phys.} {\bf G24}, 1831
  (1998).

\bibitem{Bachir:pi-K}
P.~Buettiker, S.~Descotes-Genon and B.~Moussallam, {\em Eur. Phys. J.} {\bf
  C33}, 409 (2004).

\bibitem{FOCUS}
J.~M. Link {\em et~al.}, {\em Phys. Lett.} {\bf B535}, 43 (2002).

\bibitem{Filkov:2006}
L.~V. Fil'kov and V.~L. Kashevarov, {\em Phys. Rev.} {\bf C73}, p. 035210
  (2006).

\bibitem{Mainz-pion-pol}
J.~Ahrens {\em et~al.}, {\em Eur. Phys. J.} {\bf A23}, 113 (2005).

\bibitem{Gasser:pi-pol}
J.~Gasser, M.~A. Ivanov and M.~E. Sainio, {\em Nucl. Phys.} {\bf B745}, 84
  (2006).

\bibitem{Goity:GT}
J.~L. Goity, R.~Lewis, M.~Schvellinger and L.-Z. Zhang, {\em Phys. Lett.} {\bf
  B454}, 115 (1999).

\bibitem{W:pion-scat}
S.~Weinberg, {\em Phys. Rev. Lett.} {\bf 17}, 616 (1966).

\bibitem{Gotta:atoms}
D.~Gotta, {\em Int. J. Mod. Phys.} {\bf A20}, 349 (2005).

\bibitem{AB:FW}
A.~M. Bernstein, {\em Phys. Lett.} {\bf B442}, 20 (1998).

\bibitem{BKM-photo}
V.~Bernard, N.~Kaiser and U.-G. Meissner, Eur. Phys. J., A11, 209(2001); Z.
  Phys.,C70, 483(1996); Phys. Lett.B383,116(1996) % title =.

\bibitem{SAID}
R.~A. Arndt, W.~J. Briscoe, I.~I. Strakovsky and R.~L. Workman, Phys. Rev. C74,
  045205(2006),nucl-th/0607017 , http://gwdac.phys.gwu.edu/.

\bibitem{nikos}
N.~F. Sparveris {\em et~al.}, {\em Phys. Rev. Lett.} {\bf 94}, p. 022003
  (2005).

\bibitem{Costas}
C.N.Papanicolas, Eur. Phys. J. A, 18, 141 (2003).

\bibitem{SOH}
C.~Papanicolas and A.~Bernstein, Shape of Hadrons Workshop, Athens,
  Greece(2006); http://microtron.iasa.gr/hadrons/index.html, to be published.

\bibitem{Schmidt}
A.~Schmidt {\em et~al.}, {\em Phys. Rev. Lett.} {\bf 87}, p. 232501 (2001).

\bibitem{Sask}
J.~C. Bergstrom {\em et~al.}, {\em Phys. Rev.} {\bf C53}, 1052 (1996).

\bibitem{Weis}
M.~Weis {\em et~al.}, arXiv: nucle/ex 0705.3816.

\bibitem{BKM-electro}
V.~Bernard, N.~Kaiser and U.-G. Meissner, Nucl. Phys.A607, 379(1996); Phys.
  Rev. Lett., 74, 3752(1995).

\bibitem{DMT:threshold}
S.~S. Kamalov, G.-Y. Chen, S.-N. Yang, D.~Drechsel and L.~Tiator, {\em Phys.
  Lett.} {\bf B522}, 27 (2001).

\bibitem{pi0-decay}
J.~L. Goity, A.~M. Bernstein and B.~R. Holstein, Phys. Rev.D66, 076014(2002);
  B.~Ananthanarayan and B.~Moussallam, JHEP {\bf 0205}, 052 (2002).

\bibitem{Ioffe}
B.~L. Ioffe and A.~G. Oganesian, {\em Phys. Lett.} {\bf B647}, 389 (2007).

\bibitem{COSY}
Experiments on $\eta$ and $\eta^{'}$ decays have started at COSY, H. H. Adam et
  al., nucl-ex/0411038.

\end{thebibliography}

\end{document}